\documentstyle[preprint,aps,epsfig]{revtex}
\tightenlines
\begin{document}
\draft
\preprint{\vbox{\hbox{LBNL-44607}\hbox{hep-ph/9911443}}}
\title{Anomalous Higgs-Photon Interactions in 
Photon Fusion Processes at NLC}
\author{S.\ M.\ Lietti}
\address{Theory Group, Lawrence Berkeley National Laboratory,\\
	 Berkeley, CA 94720, USA.}
\date{\today}
\maketitle
\begin{abstract}
The Standard Model (SM) of the electroweak has proven to be 
successful in describing all the available precision experimental 
data.  However, the  Higgs mechanism, responsible for the electroweak symmetry
breaking in the SM, still remains one of the most important open questions of 
the theory. The effect of new operators that give rise 
to anomalous Higgs boson coupling to two photons is examined in the 
two-photon processes $\gamma\gamma \to  H \to b\bar{b}, \gamma\gamma,
W^+W^-,ZZ$ at a high energy linear $e^+ e^-$ collider (NLC).  
\end{abstract}

\pacs{14.80.Cp}

\newpage 

\section{Introduction}
\label{introduction}

The Standard Model (SM) of the electroweak interactions based
on the gauge group $SU(2)_L \times U(1)_Y$ has proven to be 
successful in describing all the available precision experimental 
data \cite{sm_data}. This applies particularly to the predictions 
for the couplings of the gauge bosons to the matter fermions. 
The recent measurements of the gauge-boson self couplings at 
LEPII \cite{lep2} and Tevatron \cite{tevatron} collider also shed 
some light on the correctness of the SM predictions for these interactions.

On the other hand, the precise mechanism of the electroweak symmetry
breaking still remains one of the most important open questions of 
the theory. In the SM, the breaking is realized via the Higgs mechanism
in which a scalar $SU(2)$-doublet, the Higgs boson, is introduced 
{\it ad hoc} and the symmetry is spontaneously broken by the vacuum
expectation value (VEV) of the Higgs field. However, in this simple 
realization, the theory presents problems since the running
Higgs mass shows a quadratic divergence at some high scale. This may imply 
the existence of a cut-off scale $\Lambda$ above which new 
physics must appear.

The experiments which will take place at the Next Linear 
electron--positron Collider (NLC) will be able to explore 
the nature of the Higgs boson and its couplings to other 
particles \cite{review_higgs_nlc}. Deviations from the
SM predictions for these couplings would 
indicate the existence of new physics effects. 

In general, such deviations can be parametrized in terms of
effective Lagrangians by adding to the SM Lagrangian
higher dimensional operators that describe the new phenomena
\cite{effective}. This model--independent approach accounts for
new physics that shows up at an energy scale $\Lambda$, larger
than the electroweak scale. The effective Lagrangians are
constructed with the light particle spectrum that exists at low
energies, while the heavy degrees of freedom are integrated out.
They are invariant under the $SU(2)_L \times U(1)_Y$ and, in the
linearly realized version, they involve, in addition to the usual
gauge--boson fields, also the light Higgs particle. 
The most general dimension--6 effective Lagrangian, containing all SM
bosonic fields, that is  $C$ and $P$ even, was constructed in
Ref.\ \cite{hisz}.

Out of the eleven independent operators constructed in Ref.\
\cite{hisz}, three of them describe new interaction between the
Higgs particle and the photon, 
\begin{equation}
{\cal L}_{\text{eff}} =  \frac{f_{WW}}{\Lambda^2} \;
\Phi^{\dagger} \hat{W}_{\mu \nu} \hat{W}^{\mu \nu} \Phi + 
\frac{f_{BB}}{\Lambda^2} \; \Phi^{\dagger} \hat{B}_{\mu \nu}
\hat{B}^{\mu \nu} \Phi  +
\frac{f_{BW}}{\Lambda^2} \; \Phi^{\dagger} \hat{B}_{\mu \nu}
\hat{W}^{\mu \nu} \Phi\; ,
\label{lagrangian}
\end{equation}
where, in the unitary gauge, the Higgs doublet becomes $\Phi =
(1/\sqrt{2}) [\, 0 \, , \, (v + H) \, ]^T$,  $\hat{B}_{\mu
\nu} = i (g'/2) B_{\mu \nu}$, and  $\hat{W}_{\mu \nu} = i (g/2)
\sigma^a W^a_{\mu \nu}$, with $B_{\mu \nu}$ and $ W^a_{\mu \nu}$
being the field strength tensors of the $U(1)$ and $SU(2)$ gauge
fields respectively, and $\Lambda$ represents the energy scale
for new physics.

The operators of Eq. (\ref{lagrangian}) describe the effect, at
one--loop level \cite{arz}, of new heavy states predicted by the
underlying theory that should be valid at very high energies. The
possible existence of heavy fermions and/or bosons, that couple
to the (light) bosonic sector of the SM, should indirectly
manifest itself in the Higgs boson couplings via equation
(\ref{lagrangian}), after all the heavy degrees of freedom are
integrated out. Anomalous Higgs boson couplings have already been studied 
in Higgs and Z boson decays \cite{hagiwara2}, in $e^+ e^-$ \cite{ee,our,our2},
$\gamma\gamma$ \cite{gamma}, and $p\bar{p}$ colliders \cite{fer}.

In this paper, we explore the consequence of new operators that
give rise to an anomalous Higgs boson coupling to photons ($H\gamma\gamma$). 
In particular, we study the anomalous production of the Higgs boson via 
two-photon processes in a electron-positron collider, with the subsequent
decay of the Higgs boson into two particles. It is important to notice
that we have also taken into account the SM one--loop Higgs contributions 
\cite{hgg,hgz} to this vertex in our analyses. 

The lagrangian in Eq. (\ref{lagrangian}) induces, besides the $H\gamma\gamma$ 
coupling, other anomalous Higgs couplings like $HZ\gamma$, $HZZ$, and 
$HW^+W^-$. In the unitary gauge, Eq. (\ref{lagrangian}) can be written 
for the anomalous Higgs couplings as,
\begin{equation}
{\cal L}_{\text{eff}}^{\text{H}} =  
g_{H\gamma\gamma} H A_{\mu\nu} A^{\mu\nu} +
g_{HZ\gamma} H A_{\mu\nu} Z^{\mu\nu} +
g_{HZZ} H Z_{\mu\nu} Z^{\mu\nu} + 
g_{HWW} H W^+_{\mu\nu} W_-^{\mu\nu}\;\; ,
\label{lagrangian_ug}
\end{equation}
where $A_{\mu\nu} = \partial_\mu A_\nu - \partial_\nu A_\mu$ and the same for
$Z_{\mu\nu}$.
The effective couplings $g_{H\gamma\gamma}$, $g_{HZ\gamma}$, $g_{HZZ}$, 
and $g_{HWW}$ are related to the coefficients of the operators appearing
in Eq. (\ref{lagrangian}). In particular, for the $H\gamma\gamma$ coupling
one has,
\begin{equation}
g_{H\gamma\gamma} = - \frac{g m_W \sin^2\theta_W }{2}
\left(\frac{f_{BB}+f_{WW}-f_{BW}}{\Lambda^2}\right) \;\; .
\label{coupling_haa}
\end{equation}

Considering only the effect of one operator at a time and combining 
information from precision measurements at LEPI and at low energy, one has 
the following constraints at 95\% CL (in units of TeV$^{-2}$) \cite{concha}, 
for $m_H=200$ GeV and $m_{\text{top}}=175$ GeV,
\begin{eqnarray}
-1. \leq \frac{f_{BW}}{\Lambda^2} \leq 8.6 \;\;,\;\;
-79 \leq \frac{f_{BB}}{\Lambda^2} \leq 47 \;\;,\;\;
-24 \leq \frac{f_{WW}}{\Lambda^2} \leq 14 \;\;.
\label{limits}
\end{eqnarray} 

Anomalous Higgs boson production via vector boson fusion can be also
generated by the anomalous couplings $HZ\gamma$ -- $HW^+W^-$ and $HZZ$ 
of Eq. (\ref{lagrangian_ug}) are not considered here since their effect is 
much less important than the anomalous Higgs boson production via two-photon 
process due to phase space reduction.

In order to impose limits on the dimension--6 operators of Eq. 
(\ref{lagrangian}) which generate the new $H\gamma\gamma$ interaction, 
we examine the Higgs boson production via two-photon process at the 
NLC with the subsequent decay into $\gamma\gamma$, 
$b \bar{b}$, $W^+ W^-$, $ZZ$. The SM background considered for these reactions 
can be divided in two groups: 
\begin{itemize}
\item{Set I: all the SM direct contribution for 
$e^+ e^- \to \gamma\gamma$, $b \bar{b}$, $W^+ W^-$, and 
$ZZ$; }
\item{Set II: the vector boson fusion contributions 
$e^+ e^- \to W^+ W^-(\nu \bar{\nu}) \to \gamma\gamma(\nu \bar{\nu})$, 
$b \bar{b}(\nu \bar{\nu})$, $W^+ W^-(\nu \bar{\nu})$, $ZZ(\nu \bar{\nu})$, 
and $e^+ e^- \to Z Z(e^+ e^-) \to \gamma\gamma(e^+ e^-)$, 
$b \bar{b}(e^+ e^-)$, $W^+ W^-(e^+ e^-)$, $ZZ(e^+ e^-)$.}
\end{itemize}

\section{Vector Boson Fusion at NLC}

In the case of a high energy electron-positron collider, virtual gauge 
bosons can be produced nearly on--shell and collinear with the initial 
particles. If one uses the effective boson approximation \cite{eba,jot} 
one may regard the fermion beams as sources of gauge bosons and at leading 
log ignore the virtuality of these bosons in calculating the cross section. 
The process $f_a+f_b \to f_{a'}+f_{b'}+X$, where both $f_a$ and $f_b$ serve 
as source of a vector boson, can be evaluated by the effective boson 
approximation formula \cite{jot}
\begin{eqnarray}
\sigma_{f_a+f_b \to f_{a'}+f_{b'}+X}(s_0)=
\int dx_1 \int dx_2 f_n(x_1)f_m(x_2)
\hat{\sigma}^{nm}_{V_1+V_2 \to f_{a'}+f_{b'}+X}
(\hat{s}_0)
\label{evb}
\end{eqnarray}
where $\hat{s}_0=x_1 x_2 s_0$, $V_{1,2}=W^\pm$ or $Z^0$, and $m,n=-1,0,1$ 
are the vector boson 
helicities. If one writes the elementary coupling between the fermions and 
the vector boson as $\bar{\Psi} \Gamma_\mu \Psi V^\mu$ with
\begin{eqnarray}
\Gamma_\mu= g_R  \frac{\gamma_\mu(1+\gamma_5)}{2} + 
g_L  \frac{\gamma_\mu(1-\gamma_5)}{2},
\nonumber 
\end{eqnarray}
one obtains the following distribution function:
\begin{eqnarray}
f_{-1}&=&g_R^2 h_1 + g_L^2 h_2,
\nonumber \\
f_{0}&=&(g_L^2 + g_R^2) h_0,
\nonumber \\
f_{1}&=&g_L^2 h_1 + g_R^2 h_2,
\nonumber 
\end{eqnarray}
where
\begin{eqnarray}
h_0 &=& \left[ \frac{x}{16 \pi^2} \right]
\left[\frac{2(1-x)\xi}{w^2 x} - \frac{2\Delta(2-w)}{w^3} \log
\left(\frac{x}{\Delta'}\right)\right],
\nonumber \\
h_1 &=& \left[ \frac{x}{16 \pi^2} \right]
\left[\frac{-(1-x)(2-w)}{w^2}+\frac{(1-w)(\xi-w^2)}{w^3}\log
\left(\frac{1}{\Delta'}\right)
-\frac{\xi-2xw}{w^3}\log\left(\frac{1}{x}\right)\right],
\nonumber \\
h_2 &=& \left[ \frac{x}{16 \pi^2} \right]
\left[\frac{-(1-x)(2-w)}{w^2(1-w)}+\frac{\xi}{w^3}\log
\left(\frac{x}{\Delta'}\right)\right],
\nonumber 
\end{eqnarray}
where $w=x-\Delta$, $\xi=x+\Delta$, $\Delta=M_V^2/s^0$, and 
$\Delta'=\Delta/(1-w)$. 

This approach will be used to evaluate the vector boson fusion
background (Set II) described in Section \ref{introduction}.

\section{Anomalous Higgs Boson Production via Two-Photon Process at NLC}

For a high energy electron-positron collider, two-photon processes can be 
calculated using an effective photon approximation \cite{photon-aprox} so 
that if a cross section $\sigma_{\gamma \gamma \to X}$ is known, the cross 
section for $e^+e^- \to e^+e^- X$ via the two photon mechanism is given by:
\begin{equation}
\sigma_{e^+e^- \to e^+e^- X}(s_0) = \left[\frac{\alpha}{2 \pi} 
\log\left( \frac{s_0}{4 \hat{m}_e^2}\right) \right]^2 
\int_0^1 f(\tau) \sigma_{\gamma\gamma \to X}(\tau s_0) d \tau,
\label{ph-apr}
\end{equation}
where $s_0$ is the square of the center of mass energy of the initial 
$e^+e^-$ and
\begin{equation}
f(\tau)=\frac{1}{\tau}[(2+\tau)^2 \log \frac{1}{\tau} - 2(1-\tau)(3+\tau)].
\label{ftau}
\end{equation}
In this expression the total cross section for $e^+e^- \to e^+e^- X$ is given 
if one takes $\hat{m}_e = m_e = 0.5$ MeV as the mass of the electron. 
However, if one wishes to observe the $e^+e^-$ in the final state, 
experimental constrains require that a minimum cut on the transverse 
momentum of the final state electron $P_{T_{min}}$ be used. In this case 
the result is given by taking $\hat{m}_e = P_{T_{min}}$.

In order to study the anomalous Higgs boson production via two-photon process 
at NLC we do not necessarily need to observe the final $e^+ e^-$ pair because 
it is not a product of the Higgs boson decay. Besides, for a NLC with 
$\sqrt{s_0}=500(1000)$ GeV and requiring $P_{T_{min}}=20(40)$ GeV 
one finds that the cross section of Eq. (\ref{ph-apr}) is only 3.7(3.3)\% 
of the total cross section for $\hat{m}_e = m_e = 0.5$ MeV. Therefore, 
more than 96 \% of the anomalous Higgs boson production via photon fusion
happens when the final $e^+ e^-$ pair is undetected.

The anomalous contributions for the $H \gamma \gamma$ interaction are 
significant only when the Higgs boson is produced on--mass--shell, as we 
will see in section \ref{results}. For this reason, we will use the production 
of $b\bar{b}$ and $\gamma \gamma$ pairs to study a light Higgs boson mass 
range of $100 \leq m_H \leq150$ GeV at a NLC with energy $\sqrt{s_0}=500$ 
GeV and integrated luminosity ${\cal L}=50$ fb$^{-1}$. The production of 
$W^+ W^-$ and $ZZ$ pairs will be used to study a heavier Higgs boson mass 
range of $200\leq m_H\leq350$ GeV at a NLC with energy $\sqrt{s_0}=1$ TeV and 
integrated luminosity ${\cal L}=100$ fb$^{-1}$. We have considered in our 
analyses a 80\% detection efficiency for each photon and quark bottom, and 
a 80\% overall detection efficiency for the $W^+ W^-$ and $Z Z$ final state.

\section{Results}
\label{results}

In order to compute  the contributions for the signal of anomalous Higgs 
boson production via photon fusion with subsequent decay into
pairs of bottom quarks, photons, and massive gauge bosons $W$ and $Z$, 
as well as for the background for these final state pairs of particles via
direct, photon, and vector boson fusion production, we have have incorporated 
all anomalous couplings in Helas--type \cite{helas} Fortran subroutines. 
These new subroutines were used to adapt a Madgraph \cite{madgraph} output 
to include all the anomalous contributions. We have checked that our 
code passed the non--trivial test of electromagnetic gauge invariance. 
We employed Vegas \cite{vegas} to perform the Monte Carlo phase space 
integration to obtain the differential and total cross sections for the signal 
and the background (Sets I and II).

To estimate the impact of the anomalous coefficients $f_{BB}$, $f_{WW}$, and 
$f_{BW}$ in the Higgs boson production via photon fusion, we have evaluated 
the total cross section for signal and background for all processes described 
in Section (\ref{introduction}). The signal was obtained considering that all 
anomalous operators coefficients have the same value $f_{all} = f_{BB}= 
f_{BW}= f_{WW}= 30$ TeV$^{-2}$, which is in agreement with 
the limits of Eq.(\ref{limits}). In order to avoid infrared divergences in 
the two photons final state, we have required these photons to have a 
transverse momentum of $p_{T_\gamma}\geq 25$ GeV. An analysis of the 
significance of the signal (Significance = Signal/$\sqrt{\text{Background}}$) 
shows that the $b \bar{b}$ production is a better option compared to the 
$\gamma\gamma$ production in order to impose limits on the anomalous
coefficients for a light Higgs mass ($100\leq M_H(GeV) \leq 150$)
as one can see in Tables \ref{table1} and \ref{table2}. For higher 
masses ($200\leq M_H(GeV) \leq 350$), Tables \ref{table3} and \ref{table4}
show that the $W^+W^-$ production is a better option compared to the 
$ZZ$ production.

\begin{figure}
\begin{center}
\text{\epsfig{file=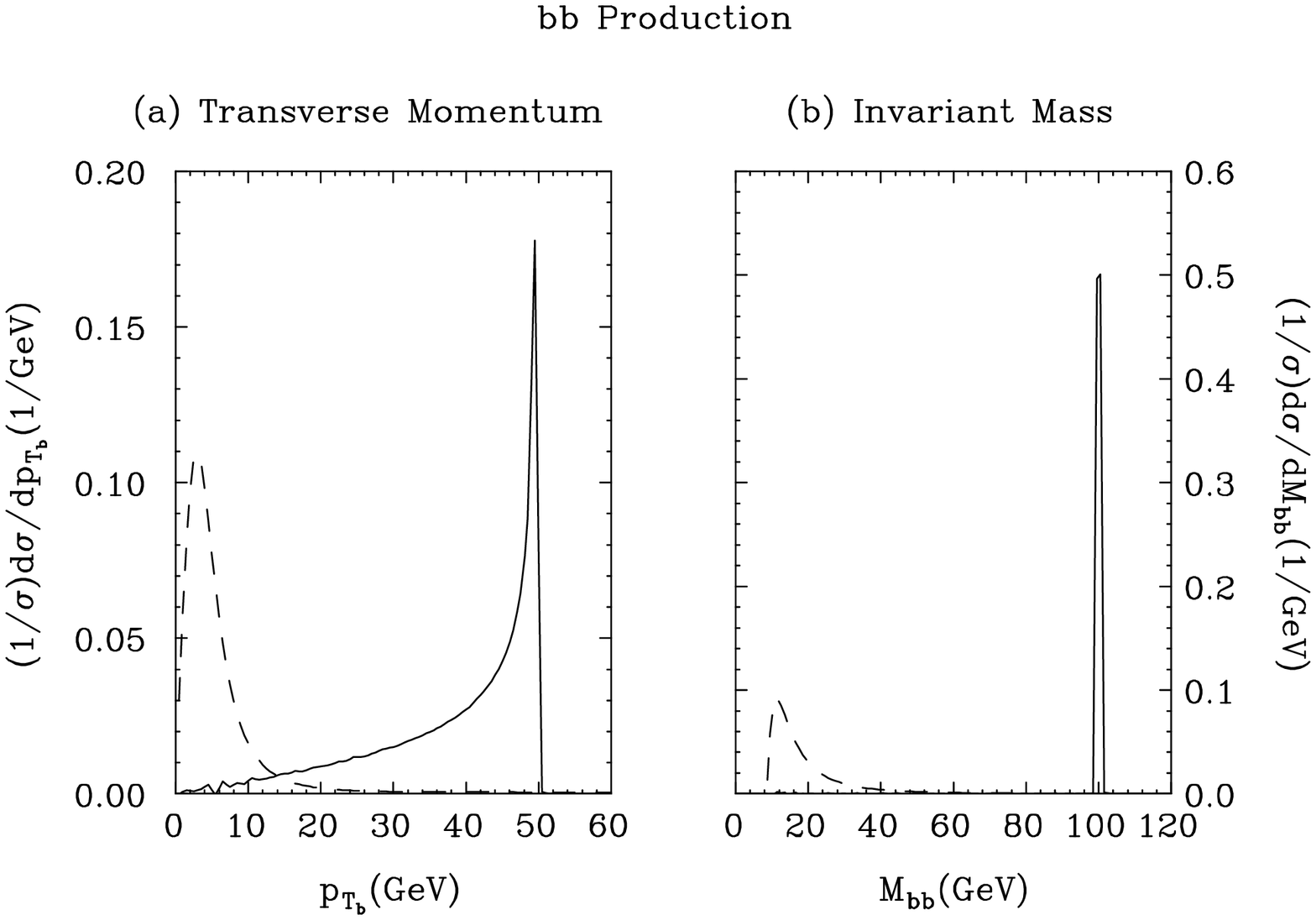,width=0.75\textwidth,angle=90}}
\end{center}
\vskip -2 cm
\caption{(a) Transverse momentum distribution of the quark bottom 
and (b) Invariant mass distribution of the $b \bar{b}$ final state particles
at NLC with $\sqrt{s_0}=500$ GeV and a Higgs boson mass of 100 GeV. 
The total background is drawn in dashed lines while the full lines are 
the signal for an anomalous Higgs production via  photon fusion 
($f_{all}=30$ TeV$^{-2}$).}
\label{fig:1}
\end{figure}

\begin{figure}
\begin{center}
\mbox{\epsfig{file=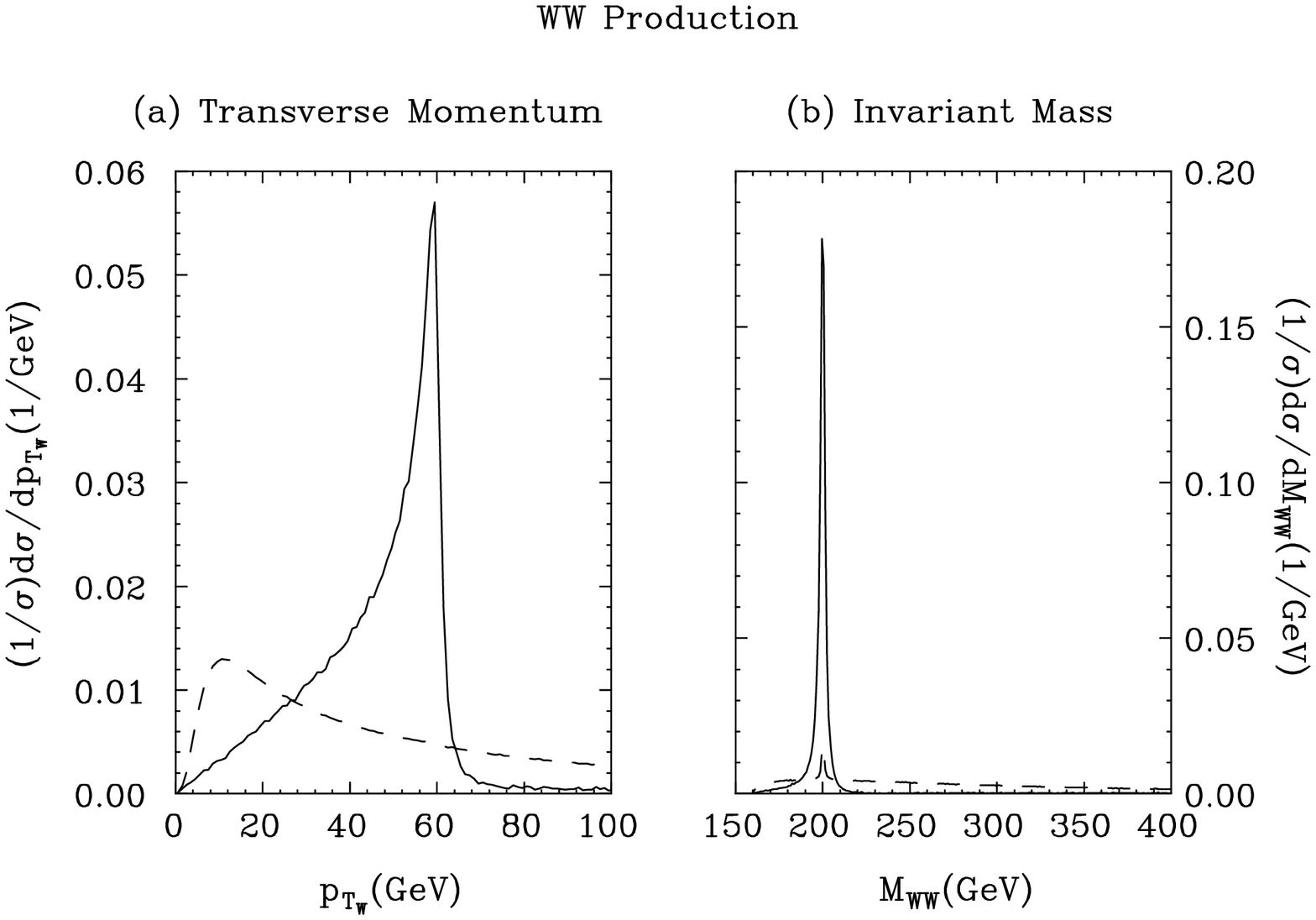,width=0.75\textwidth,angle=90}}
\end{center}
\vskip -2 cm
\caption{(a) Transverse momentum distribution of $W^+$ and (b)
Invariant mass distribution of the $W^+ W^-$ final state particles
at NLC with $\sqrt{s_0}=1$ TeV and a Higgs boson mass of 200 GeV. 
The total background is drawn in dashed lines while the full lines are 
the signal for an anomalous Higgs production via  photon fusion 
($f_{all}=30$ TeV$^{-2}$).}
\label{fig:2}
\end{figure}

\begin{figure}
\begin{center}
\mbox{\epsfig{file=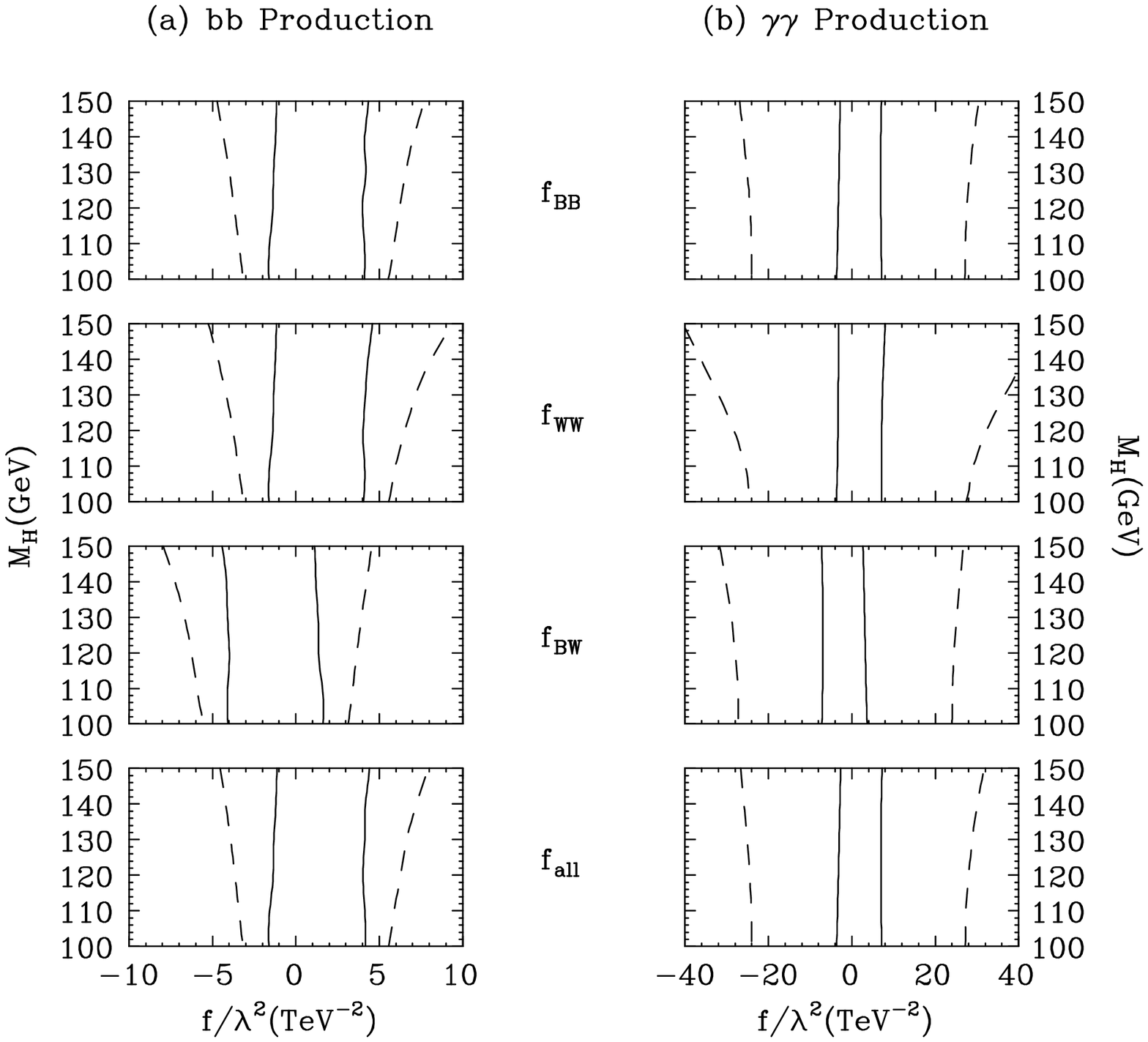,width=0.75\textwidth,angle=90}}
\end{center}
\caption{$95 \%$ CL allowed values (inside the lines) of the coefficients 
$f_{BB}$, $f_{WW}$, $f_{BW}$, and $f_{all}$, in TeV$^{-2}$, for: 
(a) $b \bar{b}$ production and (b) $\gamma \gamma$ production via photon 
fusion at NLC with $\sqrt{s}=500$ GeV and ${\cal L}=50$ fb$^{-1}$, for a 
Higgs boson mass in the range $100 \leq m_H \leq 150$ GeV. Dashed (full) 
lines are the limits obtained when a cut in the transverse momentum 
(invariant mass) distribution is used (see text).}
\label{fig:3}
\end{figure}

\begin{figure}
\begin{center}
\mbox{\epsfig{file=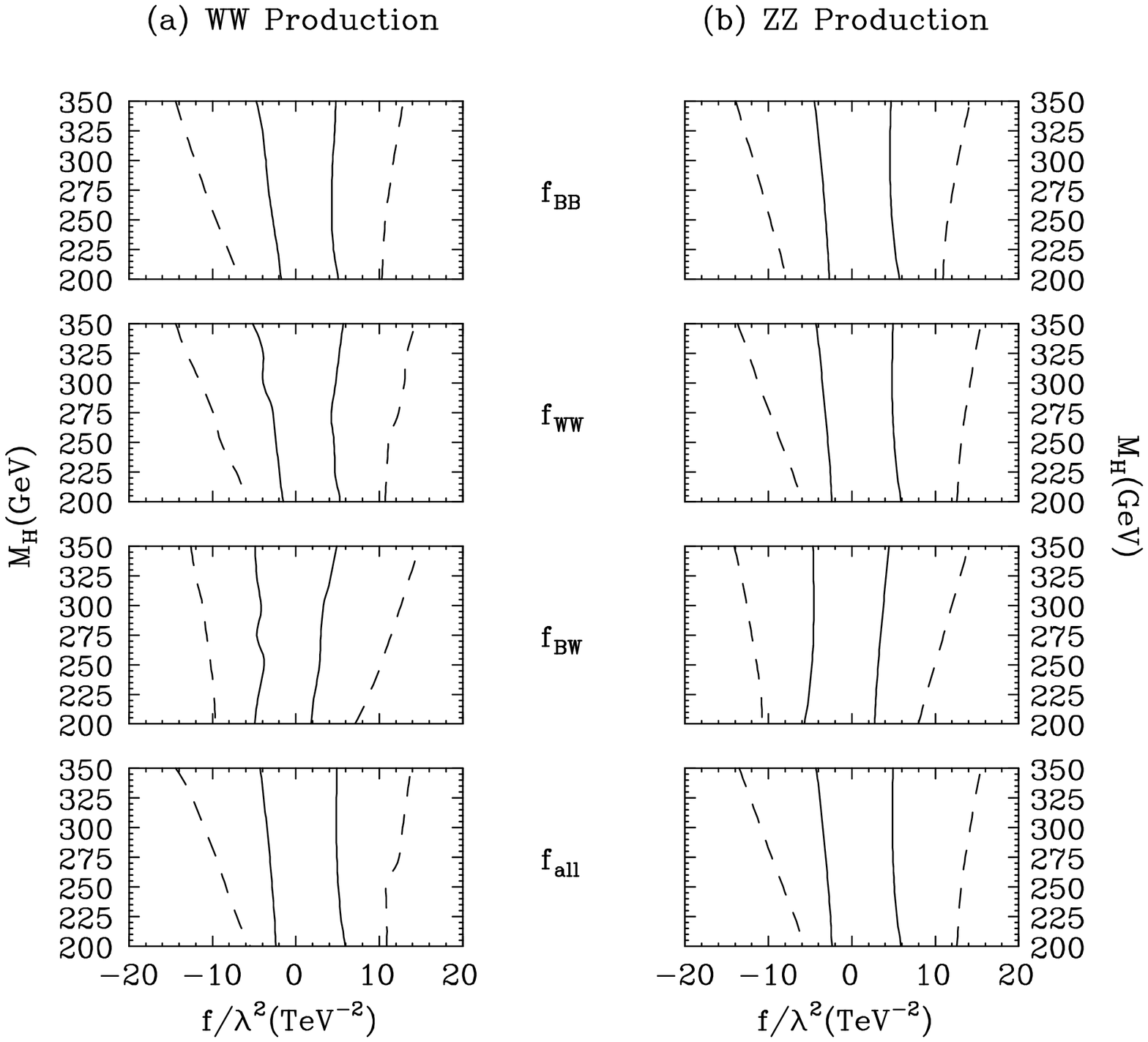,width=0.75\textwidth,angle=90}}
\end{center}
\caption{$95 \%$ CL allowed values (inside the lines) of the coefficients 
$f_{BB}$, $f_{WW}$, $f_{BW}$, and $f_{all}$, in TeV$^{-2}$, for: 
(a) $W^+ W^-$ production and (b) $Z Z$ production via photon fusion
at NLC with $\sqrt{s_0}=1$ TeV and ${\cal L}=100$ fb$^{-1}$, for a Higgs boson
mass in the range $200 \leq m_H \leq 350$ GeV. Dashed (full) lines are 
the limits obtained when a cut in the transverse momentum (invariant mass) 
distribution is used (see text).}
\label{fig:4}
\end{figure}

In order to improve the sensitivity of NLC to the anomalous Higgs boson 
production, we have investigated different distributions of the final 
state particles for both signal ($f_{all} = 30$ TeV$^{-2}$) and background. 
One of most promising variables is the transverse momentum of the final 
particles whose distribution is presented in Fig.\ \ref{fig:1} (a) for 
the $b\bar{b}$ final state and in Fig.\ \ref{fig:2} (a) for the $W^+ W^-$ 
final state. In both cases, we observe that the contribution of the anomalous 
Higgs production reaches its maximum contribution in 
\begin{equation}
p_{T_{ano}}=\frac{1}{2}\sqrt{M_H^2-M_{pair}^2},
\end{equation}
where $M_H$ is the mass of the Higgs boson and $M_{pair}$ is the sum of 
the masses of the final particles that should be a product of the Higgs 
boson decay. Similar behaviour is observed for the $\gamma \gamma$ and $ZZ$ 
final states. Therefore, we require the transverse momentum of these 
final state particles to be in the range
\begin{equation}
25  < p_T(\text{GeV}) <  (p_{T_{ano}}+5) \;\;.
\label{cut_pt}
\end{equation}
In this way, the significance of the signal is enhanced by a factor of 
at least 1.75, compared to the previous analyses without any cut, as we 
can see in Tables \ref{table1}-\ref{table4}.
The 95\% CL allowed values for the coefficients $f_{BB}$, $f_{WW}$, 
$f_{BW}$, and $f_{all}$ using the cut (\ref{cut_pt}) are shown in 
Figures \ref{fig:3} and \ref{fig:4} (dashed lines) for all the 
final pair productions. 

Another promising variable is the invariant mass of the particles 
produced in the Higgs decay, presented in Fig.\ \ref{fig:1} (b) for the 
$b\bar{b}$ final state and Fig.\ \ref{fig:2} (b) for the  $W^+ W^-$ final 
state. Since the contribution of the anomalous couplings is dominated by 
on--mass--shell Higgs production with the subsequent $H \to b \bar{b}, 
\gamma\gamma, W^+W^-,ZZ$ decays, as can be clearly seen in the 
Figures \ref{fig:1} (b) and  \ref{fig:2} (b), a more drastic cut 
would be to require 
\begin{equation}
(M_H-5) < M_{pair}^{inv}(\text{GeV}) <  (M_H+5) \;\;,
\label{cut_minv}
\end{equation}
where $M_{pair}^{inv}$ is the invariant mass of the final $b \bar{b}$,
$\gamma\gamma$,  $W^+ W^-$, or $ZZ$ pairs. The best constraints are obtained 
at NLC when this cut is applied, as we can be seen through the enhancement 
of the significance in Tables \ref{table1}-\ref{table4}. The 95\%CL results 
obtained using the cut (\ref{cut_minv}) are also shown in 
Figures \ref{fig:3} and \ref{fig:4}. These results are more restrictive than 
the constraints obtained at LEPI and at low energy [Eq. (\ref{limits})]
and for $ZZ\gamma$ and $Z\gamma\gamma$ production at LEPII and NLC 
\cite{our2}, especially for $f_{BB}$ and $f_{WW}$.

\section{Conclusions}

The search for the effect of higher dimensional operators that give rise to 
anomalous Higgs boson couplings may provide important information on physics 
beyond the SM and should be pursued in all possible reactions. In this paper, 
we have studied the $b \bar{b}$, $\gamma\gamma$, $W^+ W^-$,and $ZZ$ 
production in high energy $e^+ e^-$ colliders (NLC) via photon fusion, 
focusing on the operators that generate anomalous $H\gamma\gamma$ coupling. 

We established the limits that can be imposed at NLC through the analysis of 
the impact of the anomalous coupling over the total cross section of processes 
involving two final bottons, photons, W's, and Z's bosons. In order to improve 
the sensitivity of NLC to this anomalous Higgs boson production, the limits 
were evaluated for the cases where a convenient cut on the transverse 
momentum spectrum and on the invariant mass spectrum of the final state 
particles is used. 

Typical values of a few TeV$^{-2}$ are reached in our analyses. Our results 
are more restrictive than the constraints obtained at low energy data, 
at LEPI, and for $ZZ\gamma$ and $Z\gamma\gamma$ production at LEPII and NLC 
\cite{our2}. Therefore, the NLC should provide important hints about the 
existence of new physics beyond the Standard Model. 

\acknowledgments
I would like to thank S.F.~Novaes and J.K.~Mizukoshi for very useful 
discussions. This work was supported in part by the Director,
Office of Science, Office of Science, Office of Basic Energy Services, 
of the U.S. Department of Energy under Contract DE-AC03-76SF00098 and in 
part by Funda\c{c}\~ao de Amparo \`a Pesquisa do Estado de S\~ao Paulo 
(FAPESP).


\begin{table}
\begin{tabular}{||c||c||c||c||}
Cross Section (fb)& Without Cuts & 
Transverse Momentum Cut &
Invariant Mass Cut \\
\hline 
Background ($e^+ e^- \to b \bar{b}$) &381 & 10.9 & 0 \\
\hline
Background ($\gamma \gamma \to b \bar{b}$) & 783 &6.8 &1.7 \\
\hline
Background ($W^+ W^- \to b \bar{b}$) &9.0 & 6.2 & 3.1 \\
\hline
Background ($Z Z \to b \bar{b}$) & 0.36 & 0.16 & 0.13 \\
\hline
Signal ($\gamma \gamma \to H \to b \bar{b}$) & 66.4 & 57.6 &66.4 \\
\hline
\hline
Significance & 11 & 66 & 168
\end{tabular}
\caption{Background and  Signal ($f_{all}=30$ TeV$^{-2}$ and 
$M_H=100$ GeV) cross sections (in fb) for the $b \bar{b}$ final state 
at NLC with a center of mass  energy of 500 GeV. The Significance is given by
the fraction $\left(\frac{\text{Number of Signal events}}
{\sqrt{\text{Number of Background events}}}\right)$ for 
${\cal L}=50$ fb$^{-1}$ and $e_f$=64\% overall detection efficiency.
The Transverse Momentum Cut is 
$25<P_{T_{b,\bar{b}}}(GeV)<(P_{T_{ano}}+5)$ while the Invariant Mass
Cut is $(M_H-5)< M^{inv}_{b \bar{b}}(GeV)<(M_H+5)$.}
\label{table1}
\end{table}

\begin{table}
\begin{tabular}{||c||c||c||c||}
Cross Section (fb) & $P_{T_{\gamma_1,\gamma_2}}>25$ GeV & 
Transverse Momentum Cut &
Invariant Mass Cut\\ 
\hline 
Background ($e^+ e^- \to \gamma \gamma$) &2981 & 942 & 0 \\
\hline
Background ($\gamma \gamma \to \gamma \gamma$) & $< 8 \times 10^{-5}$ &
$< 8 \times 10^{-5}$ & $< 8 \times 10^{-5}$ \\
\hline
Background ($W^+ W^- \to \gamma \gamma$) & $< 3 \times 10^{-2}$ &
$< 2 \times 10^{-2}$ & $< 5 \times 10^{-3}$ \\
\hline
Background ($Z Z \to \gamma \gamma$) & $<9 \times 10^{-5}$ &
$<9 \times 10^{-5}$ & $< 9 \times 10^{-5}$ \\
\hline
Signal ($\gamma \gamma \to H \to \gamma \gamma$) &14.7&12.7&14.7\\
\hline
\hline
Significance & 1.5 & 2.3 & $>$1200
\end{tabular}
\caption{Background and  Signal ($f_{all}=30$ TeV$^{-2}$ and 
$M_H=100$ GeV) cross sections (in fb) for the $\gamma\gamma$ final state 
at NLC with a center of mass  energy of 500 GeV. The Significance is given by
the fraction $\left(\frac{\text{Number of Signal events}}
{\sqrt{\text{Number of Background events}}}\right)$ for 
${\cal L}=50$ fb$^{-1}$ and $e_f$=64\% overall detection efficiency.
The Transverse Momentum Cut is 
$25<P_{T_{\gamma_1,\gamma_2}}(GeV)<(P_{T_{ano}}+5)$ while the Invariant Mass
Cut is $(M_H-5)< M^{inv}_{\gamma\gamma}(GeV)<(M_H+5)$.}
\label{table2}
\end{table}

\newpage

\begin{table}
\begin{tabular}{||c||c||c||c||}
Cross Section (fb)& Without Cuts & 
Transverse Momentum Cut &
Invariant Mass Cut \\
\hline 
\hline
Background ($e^+ e^- \to W W$) & 2655 & 633 &0 \\
\hline
Background ($\gamma \gamma \to W W$) & 196 & 87 &9.8  \\
\hline
Background ($W^+ W^- \to W W$)& 5.4 & 3.4 & 2.5 \\
\hline
Background ($Z Z \to W W$) & 0.42 & 0.32 &0.27 \\
\hline
Signal ($\gamma \gamma \to H \to W^+ W^-$)& 36.3 & 33.8 & 31.9 \\
\hline
\hline
Significance & 6 & 11 & 81
\end{tabular}
\caption{Background and Signal ($f_{all}=30$ TeV$^{-2}$ and 
$M_H=200$ GeV) cross sections (in fb) for the $W^+ W^-$ final state 
at NLC with a center of mass  energy of 1 TeV. The Significance is given by
the fraction $\left(\frac{\text{Number of Signal events}}
{\sqrt{\text{Number of Background events}}}\right)$ for 
${\cal L}=100$ fb$^{-1}$ and $e_f$=80\% overall detection efficiency.
The Transverse Momentum Cut is 
$25<P_{T_{W^+, W^-}}(GeV)<(P_{T_{ano}}+5)$ while the Invariant Mass
Cut is $(M_H-5)< M^{inv}_{W^+ W^-}(GeV)<(M_H+5)$.}
\label{table3}

\end{table}
\begin{table}
\begin{tabular}{||c||c||c||c||}
Cross Section (fb)& Without Cuts & 
Transverse Momentum Cut &
Invariant Mass Cut \\
\hline 
\hline
B($e^+ e^- \to Z Z$)& 147 & 21.2 &0 \\
\hline
B($\gamma \gamma \to Z Z$)& 0.06 & 0.05 &0.05 \\
\hline
B($W^+ W^- \to Z Z$)& 1.58 & 0.88 & 0.92 \\
\hline
B($Z Z \to Z Z$) &0.10 & 0.08 &0.09 \\
\hline
S($\gamma \gamma \to H \to Z Z$)& 6.9 & 4.6 & 5.2\\
\hline
\hline
Significance ($S/\sqrt{B}$) & 5 & 9 & 45
\end{tabular}
\caption{Background and  Signal ($f_{all}=30$ TeV$^{-2}$ and 
$M_H=200$ GeV) cross sections (in fb) for the $ZZ$ final state 
at NLC with a center of mass  energy of 1 TeV. The Significance is given by
the fraction $\left(\frac{\text{Number of Signal events}}
{\sqrt{\text{Number of Background events}}}\right)$ for 
${\cal L}=100$ fb$^{-1}$ and $e_f$=80\% overall detection efficiency.
The Transverse Momentum Cut is 
$25<P_{T_{Z_1, Z_2}}(GeV)<(P_{T_{ano}}+5)$ while the Invariant Mass
Cut is $(M_H-5)< M^{inv}_{ZZ}(GeV)<(M_H+5)$.}
\label{table4}
\end{table}

\end{document}